\documentclass[%
 reprint,
 superscriptaddress,
nofootinbib,
 amsmath,amssymb,
 aps,
prd,
showkeys,
]{revtex4-2}

\usepackage{graphicx}
\usepackage{dcolumn}
\usepackage{bm}
\usepackage{cancel}

\usepackage{soul}

\usepackage{natbib}
\usepackage{graphicx,amsfonts,amssymb,amsbsy}
\usepackage{amsmath,amsthm,latexsym}
\usepackage{mathrsfs}
\usepackage[utf8]{inputenc} 
\usepackage{tabularx} 
\usepackage{multirow} 

\usepackage[normalem]{ulem}
\usepackage{fancyvrb}
\usepackage{fancyhdr}
\usepackage{placeins}
\usepackage[colorlinks]{hyperref}

\hypersetup{
    citecolor=blue,  
    linkcolor=red,  
    urlcolor=blue    
}

\usepackage{comment}

\begin{document}

\preprint{APS/123-QED}

\title{The Delta-isobar masquerade: intrahadronic phase transitions and their quark-mimicking signatures in neutron stars}

\author{Martin O. Canullan-Pascual}
 \email{canullanmartin@fcaglp.unlp.edu.ar}
 \homepage{https://arco.fcaglp.unlp.edu.ar}
 \affiliation{Grupo de Astrofísica de Remanentes Compactos\\ Facultad de Ciencias Astronómicas y Geofísicas, Universidad Nacional de La Plata\\ Paseo del Bosque S/N, La Plata (1900), Argentina.\\}
 \affiliation{CONICET, Godoy Cruz 2290, Buenos Aires (1425), Argentina}
 
\author{Germán Lugones}
 \email{german.lugones@ufabc.edu.br}
\affiliation{Universidade Federal do ABC, Centro de Ciências Naturais e Humanas,\\ Avenida dos Estados 5001- Bangú, CEP 09210-580, Santo André, SP, Brazil.}

\author{Ignacio F. Ranea-Sandoval}
\affiliation{Grupo de Astrofísica de Remanentes Compactos\\ Facultad de Ciencias Astronómicas y Geofísicas, Universidad Nacional de La Plata\\ Paseo del Bosque S/N, La Plata (1900), Argentina.\\}
 \affiliation{CONICET, Godoy Cruz 2290, Buenos Aires (1425), Argentina}

\author{Milva G. Orsaria}
\affiliation{Grupo de Astrofísica de Remanentes Compactos\\ Facultad de Ciencias Astronómicas y Geofísicas, Universidad Nacional de La Plata\\ Paseo del Bosque S/N, La Plata (1900), Argentina.\\}
 \affiliation{CONICET, Godoy Cruz 2290, Buenos Aires (1425), Argentina}%

\date{\today}

\begin{abstract}
We investigate the conditions under which $\Delta(1232)$ isobars trigger a first-order phase transition within purely hadronic neutron-star matter, using the SW4L relativistic mean-field parametrization. For scalar–vector coupling differences $0.15 \lesssim x_{\sigma\Delta} - x_{\omega\Delta} \lesssim 0.2$ and $x_{\sigma\Delta} \gtrsim 1.3$, the onset of $\Delta^-$ resonances produces a van der Waals–like instability driven by a self-amplifying feedback in the scalar meson sector, in which the $\Delta^-$ particle fraction acts as the order parameter of a Landau-type transition. A Maxwell construction yields a sharp density discontinuity at baryon densities $n_b \sim (1.3$–$2)\,n_0$, separating a $\Delta$-free outer core from a $\Delta$-rich inner core. The resulting neutron-star sequences satisfy all current multimessenger constraints: maximum masses $M_{\rm max} \approx 2.15$–$2.25\,M_\odot$, radii $R_{1.4} \approx 11$–$12$~km, and tidal deformabilities $\Lambda_{1.4} \approx 190$–$480$, compatible with NICER observations and GW170817. We compute, for the first time for a $\Delta$-induced interface, the $\ell = 2$ composition $g$-mode eigenfrequencies, obtaining $\nu_g \sim 400$–$1100$~Hz with gravitational-wave damping times $\tau_g \sim 10^3$–$10^9$~s. These frequencies overlap quantitatively with those predicted for hadron–quark phase-transition interfaces, demonstrating that the mass–radius ``knee'', reduced tidal deformability, and $g$-mode spectrum conventionally regarded as signatures of quark deconfinement can be reproduced by a purely intrahadronic mechanism. This extends the masquerade problem from static observables to the domain of gravitational-wave asteroseismology, implying that a future detection of a discontinuity $g$-mode alone would not suffice to identify quark matter in neutron-star cores. Our models are broadly consistent with recent model-agnostic Bayesian constraints on first-order phase transitions, and we show that microphysics-informed correlations among transition properties can populate regions of the equation-of-state space that are undersampled by generic parametrizations.
\end{abstract}

\keywords{equation of state; dense matter; stars: neutron}
\maketitle


\section{Introduction}

Phase transitions in dense matter can dramatically alter the internal structure and observable properties of neutron stars (NSs).  In the context of quantum chromodynamics (QCD), it is widely expected that hadronic matter may undergo a deconfinement transition to quark matter at supranuclear densities, but the relevant density range and the corresponding microphysics remain insufficiently constrained by first principles and current observations.  By contrast, purely \emph{intrahadronic} first-order phase transitions (FOPT) within the confined sector have received comparatively less attention from a dynamical and observational perspective.  While earlier work explored FOPTs associated with meson condensation (pion and kaon condensates~\cite{Haensel:1982pci,Glendenning:1999fok}), we focus here on a qualitatively different intrahadronic mechanism: the onset of $\Delta(1232)$ resonances.

Many studies have shown that, depending on their meson couplings,
$\Delta$-isobars in $\beta$-equilibrated matter may appear already at
densities of a few times nuclear saturation and can substantially soften
the equation of state
(EoS)~\cite{Cai:2015cda,Zhu:2016dri,Sahoo:2018nsm}.  This raises a
well-known tension---often referred to as the \emph{$\Delta$-isobar
puzzle}---between an early $\Delta$ onset (which generally favors
compact stars with smaller radii) and the existence of massive NSs with
masses around $2\,M_\odot$ (which require sufficient stiffness at high
density)~\cite{Drago:2014ita,Li:2018cbd,Ribes:2019ibd}.  In most
existing treatments within the relativistic mean-field (RMF) framework,
the $\Delta$ population turns on smoothly, acting as an additional
degree of freedom that gradually reshapes the composition and stiffness
of the EoS without introducing
instabilities~\cite{Malfatti:2020dba}.

For a particular window of meson--$\Delta$ couplings, however, the
onset of $\Delta$ resonances is not smooth but triggers a genuine
FOPT.  De~Oliveira
\textit{et~al.}~\cite{DeOliveira:2007eod,Deoliveira:2016pti} first
reported a negative compressibility ($dP/dn_b < 0$) and an associated
Maxwell construction in a nonlinear Walecka model when the
$\Delta$--$\omega$ coupling is sufficiently reduced relative to the
$\Delta$--$\sigma$ coupling.  Lavagno and
Pigato~\cite{Lavagno:2019tii} analyzed the mechanical and
chemical-diffusive instabilities that arise in the SFHo model with
hyperons and $\Delta$-isobars, applying a Gibbs construction; however,
the resulting maximum masses ($M_{\max}\lesssim 1.1\,M_\odot$) are
incompatible with current pulsar constraints.
Raduta~\cite{Raduta:2021dan} carried out the most comprehensive study
to date, mapping the spinodal region in the DDME2 covariant density
functional and showing that an outer-core instability at
\mbox{$n_b \sim (1$--$2)\,n_0$} driven by the $\Delta^-$ is the only scenario
consistent with both the Maxwell construction and
$M_{\max}\gtrsim 2\,M_\odot$; this analysis also quantified the impact
on tidal deformabilities ($\Lambda$) and on the direct Urca threshold.

Despite this body of work at the EoS level, the consequences of a
$\Delta$-driven density discontinuity for the neutron-star oscillation
spectrum---in particular, the discontinuity $g$-mode supported by the
sharp interface---have not been explored.

Independently of the specific microphysics, model-agnostic Bayesian analyses have recently begun to constrain the occurrence and strength of FOPTs along the neutron-star EoS~\cite{Gorda:2022lsk,Brandes:2023bob,Essick:2023fso,Komoltsev:2024lcr}. Gorda \textit{et~al.}~\cite{Gorda:2022lsk} mapped the allowed parameter space of transition onset density and latent energy using piecewise polytropes anchored to chiral EFT and perturbative QCD, while Brandes \textit{et~al.}~\cite{Brandes:2023bob} inferred the speed of sound through a Bayes-factor analysis incorporating the heavy ($2.35\,M_\odot$) black-widow pulsar PSR~J0952--0607. Essick \textit{et~al.}~\cite{Essick:2023fso} introduced a complementary nonparametric approach that extracts phase-transition features directly from macroscopic observables without relying on an underlying EoS parametrization, and found that current data disfavor only the strongest transitions (latent energy per particle $\Delta(E/N) \gtrsim 100$~MeV).  While these studies conclude that strong FOPTs are disfavored within their respective frameworks, as noted by Komoltsev~\cite{Komoltsev:2024lcr}, the parametric approaches of Refs.~\cite{Gorda:2022lsk,Brandes:2023bob} do not explicitly model the FOPT itself.  In a fully Bayesian inference with explicit inclusion of FOPTs using Gaussian processes, Komoltsev~\cite{Komoltsev:2024lcr} found that the current data \emph{cannot} distinguish between a smooth crossover and a FOPT: the Bayes factors between the two scenarios are of order unity, and roughly $91\%$ of the total evidence consists of EoSs exhibiting some form of phase change.  These complementary approaches---and the sensitivity of their conclusions to parametrization and prior choices---motivate a direct comparison with concrete microscopic models, which we undertake below for the $\Delta$-driven transition.

The absence of $g$-mode calculations for $\Delta$-driven density
discontinuities is particularly significant in light of the ongoing
effort to identify the composition of neutron-star cores.  Although
masses, radii, and tidal deformabilities provide powerful constraints
on the EoS, they are not, by themselves, sufficient to distinguish a
hybrid star containing deconfined quark matter from a purely hadronic
one---a degeneracy known as the \emph{masquerade}
problem~\cite{Alford:2004hst}.  The presence of a density gap
introduces features that are, in principle, more discriminating:
``knees'' in mass--radius ($M$--$R$) relations and interface $g$-modes
are conventionally regarded as distinctive signatures of a
deconfinement phase transition to quark
matter~\cite{Finn:1987gmi,Miniutti:2002nro,Sotani:2001ddo,Flores:2013dha,Ranea:2018omo,Tonetto:2020dgm,Rodriguez:2021hsw,Rodriguez:2025cog},
precisely because purely hadronic models with a smooth composition
change do not produce them.  As we demonstrate below, however, the
$\Delta$-induced hadron--hadron phase transition generates both a
pronounced $M$--$R$ knee and $g$-mode frequencies in the same range as
those predicted for quark-hybrid stars, undermining the discriminating
power of these observables and extending the masquerade into the domain
of gravitational-wave asteroseismology.

We implement this program within the SW4L RMF
parametrization~\cite{Spinella:2019hns,Malfatti:2020dba,Celi:2024doh}
and perform a targeted scan of the meson--$\Delta$ coupling ratios.
Confirming and extending the findings of
Refs.~\cite{DeOliveira:2007eod,Lavagno:2019tii,Raduta:2021dan} in a
different hadronic framework, we show that for
$x_{\sigma\Delta}\gtrsim 1.3$ and
$0.15 \lesssim x_{\sigma\Delta}-x_{\omega\Delta}\lesssim 0.2$,
$\Delta$ formation triggers a  hadron--hadron FOPT at densities $n_b\sim (1.5$--$2)\,n_0$.  Beyond this
confirmation, our analysis contributes three elements that are absent
from the existing literature.
\emph{(i)}~We trace the origin of the instability to a self-amplifying
feedback in the scalar meson sector and provide a Landau-type
interpretation in which the $\Delta^-$ fraction serves as the order
parameter, clarifying why the transition is first-order for
$\Delta x \gtrsim 0.15$ and becomes a smooth crossover for smaller
coupling differences.
\emph{(ii)}~We compute, for the first time for a $\Delta$-induced
interface, the $\ell=2$ non-radial composition $g$-mode
eigenfrequencies and gravitational-wave damping times, obtaining
$\nu_g \sim 400$--$1100$~Hz and \mbox{$\tau_g \sim 10^3$--$10^9$~s}.
\emph{(iii)}~We show that these $g$-mode frequencies overlap
quantitatively with those predicted for hadron--quark phase-transition
interfaces~\cite{Tonetto:2020dgm,Rodriguez:2021hsw,Rodriguez:2025cog},
extending the masquerade problem~\cite{Alford:2004hst} from static
observables ($M$, $R$, $\Lambda$) to the dynamical oscillation spectrum.
This result underscores that a future detection of a discontinuity
$g$-mode alone would not suffice to identify quark deconfinement.

This paper is organized as follows.  In Sec.~\ref{sec:eos} we summarize
the RMF framework and the treatment of $\Delta$ isobars.  In
Sec.~\ref{sec:results} we present our results: we first analyze the
thermodynamics and microphysics of the $\Delta$-induced hadron--hadron
phase transition, then discuss its impact on the $M$--$R$ relation,
tidal deformability, and nonradial $g$-modes, and finally compare
our EoS models with recent model-agnostic Bayesian constraints.
We conclude in Sec.~\ref{sec:summary} with a discussion of
implications and outlook.

\section{Theoretical Framework}
\label{sec:eos}

To describe strongly interacting matter in the NS core, we adopt a Relativistic Mean Field (RMF) framework. Specifically, we employ the SW4L parametrization \cite{Spinella:2019hns,Malfatti:2020dba,Celi:2024doh}, a density-dependent hadronic model calibrated to nuclear and hypernuclear phenomenology. For the crustal region, we smoothly join this core EoS to the SCVBB model \cite{Sharma:2015ueo} at lower densities.

In the present study, we extend the standard SW4L hadronic setup to investigate the effects of $\Delta$-resonances. We focus exclusively on the hadronic sector to isolate the conditions under which a hadron--hadron FOPT, between a $\Delta$-free and a $\Delta$-rich phase, may be triggered by the onset of $\Delta$ degrees of freedom.

\subsection{The Lagrangian Density}

The total Lagrangian density describes the interaction between baryons via the exchange of meson fields, supplemented by a free leptonic sector. The interactions are mediated by scalar-isoscalar ($\sigma, \sigma^*$), vector-isoscalar ($\omega, \phi$), and vector-isovector ($\rho$) mesons. The full Lagrangian density is given by
\begin{equation}
\mathcal{L} =  \sum_B \mathcal{L}_B + \mathcal{L}_\sigma  + \mathcal{L}_{\sigma^*} + \mathcal{L}_\omega + \mathcal{L}_\phi + \mathcal{L}_\rho +  \sum_L \mathcal{L}_L \,,
\label{eq:full_Lagrangian}
\end{equation}
where the sum over $B$ runs over the nucleons ($p, n$), the $\Lambda, \Sigma, \Xi$ hyperons, and the four $\Delta(1232)$ states ($\Delta^{++}, \Delta^+, \Delta^0, \Delta^-$). The sum over $L$ runs over the leptons $e^-$ and $\mu^-$.

The contribution of a generic baryon species $B$ (with Dirac field $\psi_B$ and mass $m_B$) reads
\begin{equation}
\begin{split}
\mathcal{L}_B 
= \bar{\psi}_B \Big[ \gamma_{\mu}  ( i\partial^{\mu} 
&- g_{\omega B}   \omega^{\mu}   - g_{\phi B}   \phi^{\mu}  
- \tfrac{1}{2} g_{\rho B}(n_b)   \boldsymbol{\tau} \cdot \boldsymbol{\rho}^{\mu} ) \\
&- ( m_B - g_{\sigma B}  \sigma - g_{\sigma^* B}  \sigma^*) \Big] \psi_B \,,
\end{split}
\label{Eq:laghad}
\end{equation}
where $n_b$ is the total baryon number density, and $g_{\rho B}(n_b)$ is the density-dependent isovector coupling defined below. The Pauli matrices $\boldsymbol{\tau}$ act in isospin space.

The mesonic and leptonic contributions entering Eq.~\eqref{eq:full_Lagrangian} are given by
\begin{align}
\mathcal{L}_\sigma &= \tfrac{1}{2} \big( \partial_{\mu}\sigma\,\partial^{\mu}\sigma - m^2_{\sigma}\sigma^2 \big) - U(\sigma)\,, \\
\mathcal{L}_{\sigma^*} &= \tfrac{1}{2} \big( \partial_{\mu}\sigma^*\,\partial^{\mu}\sigma^* - m^2_{\sigma^*}\sigma^{*2} \big)\,, \\
\mathcal{L}_\omega &= - \tfrac{1}{4}\,\omega_{\mu\nu}\omega^{\mu\nu} + \tfrac{1}{2} m^2_{\omega}\,\omega_{\mu}\omega^{\mu}\,, \\
\mathcal{L}_\phi  &= - \tfrac{1}{4}\,\phi_{\mu\nu}\phi^{\mu\nu}   + \tfrac{1}{2} m^2_{\phi}\,\phi_{\mu}\phi^{\mu}\,, \\
\mathcal{L}_\rho  &= - \tfrac{1}{4}\,\boldsymbol{\rho}_{\mu\nu} \cdot \boldsymbol{\rho}^{\mu\nu} 
                    + \tfrac{1}{2} m^2_{\rho}\,\boldsymbol{\rho}_{\mu} \cdot \boldsymbol{\rho}^{\mu}\,, \\
\mathcal{L}_L     &= \bar{\psi}_L \big( i \gamma_\mu \partial^\mu - m_L \big) \psi_L \,.
\end{align}
where the scalar self-interaction potential is defined as
\begin{equation}
U(\sigma) =  \tfrac{1}{3} b_{\sigma}\,m_N \big(g_{\sigma N} \sigma\big)^3 
          + \tfrac{1}{4} c_{\sigma} \big(g_{\sigma N} \sigma\big)^4 \, .
\end{equation}
The meson field-strength tensors are defined in the usual way:
\begin{align}
\omega_{\mu\nu} &= \partial_\mu \omega_\nu - \partial_\nu \omega_\mu\,, \\
\phi_{\mu\nu}   &= \partial_\mu \phi_\nu   - \partial_\nu \phi_\mu\,,   \\
\boldsymbol{\rho}_{\mu\nu} &= \partial_\mu \boldsymbol{\rho}_\nu - \partial_\nu \boldsymbol{\rho}_\mu\,.
\end{align}

\subsection{Parametrization and $\Delta$-couplings}

The meson-baryon coupling constants determine the stiffness and composition of the EoS. In the SW4L parametrization, the couplings of the nucleon and hyperon sectors are calibrated based on the Nijmegen extended-soft-core (ESC08) model and nuclear phenomenology \citep{Spinella:2019hns, Malfatti:2020dba,Celi:2024doh}. A key feature of this model is the density dependence of the isovector coupling, $g_{\rho B}(n_b)$, given by:
\begin{equation}
g_{\rho B}(n_b) = g_{\rho B}(n_0) \exp \left[ -a_{\rho} \left( \frac{n_b}{n_0} - 1 \right) \right],
\end{equation}
where $n_b = \sum_B n_B$ is the total baryon density, \mbox{$n_0 = 0.15$ fm$^{-3}$} is the saturation density, and \mbox{$a_{\rho} = 4.06 \times 10^{-3}$}. This dependence allows for a better description of the symmetry energy slope and neutron skin thickness.

While the nucleon and hyperon couplings are constrained by experimental data and symmetry arguments, the interaction strength of $\Delta$-resonances remains poorly determined. The couplings are typically expressed as ratios to the nucleon couplings, $x_{i\Delta} \equiv g_{i\Delta}/g_{iN}$.  Following the analysis of Ref. \cite{canullan:2025nss}, we explore the parameter space where the appearance of $\Delta$ resonances leads to distinct macroscopic behaviors. Specifically, we 
investigate extensions of the $\Delta$--$\sigma$ coupling, $x_{\sigma\Delta}$, beyond  the standard  values found in the literature, where the intrahadronic phase transitions do not occur (see, for example Refs. \cite{Malfatti:2020dba, Kalita:2024pti} and references therein). Variations in $x_{\sigma\Delta}$ directly affect the effective mass of the baryons and, consequently, the onset density of the resonances. Table~\ref{tab:eos_params} summarizes the specific sets of coupling ratios employed in this work. For the $\rho-\Delta$ interaction, we adopt universal coupling, i.e., $x_{\rho\Delta} = 1$.

\begin{table*}[t]
\centering
\caption{Model parameters and properties of the hadron-hadron phase transition. We list the meson-$\Delta$ coupling ratios ($x_{\sigma\Delta}, x_{\omega\Delta}$) used to generate the four EoS studied in this work. The columns $n_{\text{onset}}$ and $n_{\text{end}}$ indicate the baryon number densities (in units of saturation density $n_0$) at the beginning and at the end of the density jump associated with the phase transition interface. The transition pressure is $P_{\rm tr}$ and $\Delta \varepsilon$ is the energy density jump. The quantity $\varepsilon_{\rm onset}$ denotes the energy density
at which the phase transition begins and is used in the evaluation of the Seidov stability criterion discussed in Sec.~\ref{sec:results_B}.}
\label{tab:eos_params}
\begin{ruledtabular}
\begin{tabular}{ccccccccc}
EoS & $x_{\sigma\Delta}$ & $x_{\omega\Delta}$ & $\Delta x \equiv x_{\sigma\Delta} - x_{\omega\Delta}$ & $n_{\text{onset}} (n_0)$ & $n_{\text{end}} (n_0)$ & $P_{\rm tr}$ (MeV fm$^{-3}$)& $\varepsilon_{\rm onset}$ (MeV fm$^{-3}$)
 & $\Delta \varepsilon$ (MeV fm$^{-3}$) \\
\hline
eos-1 & 1.45 & 1.25 & 0.20 & 1.32 & 1.96 & 6.48 & 202.25 & 101.47 \\
eos-2 & 1.45 & 1.30 & 0.15 & 1.48 & 1.82 & 9.54 & 228.32 & 54.92 \\
eos-3 & 1.30 & 1.10 & 0.20 & 1.41 & 2.04 & 8.21 & 217.55 & 100.73 \\
eos-4 & 1.30 & 1.15 & 0.15 & 1.61 & 1.78 & 11.53& 249.52 & 27.22 \\
\end{tabular}
\end{ruledtabular}
\end{table*}

\subsection{Thermodynamics and Equations of Motion}

The equations of motion for the meson fields are derived in the RMF approximation, where bars denote the mean values of the time-like components of the meson fields in uniform matter (e.g., $\bar{\omega} \equiv \langle \omega^0 \rangle$). The coupled non-linear equations are
\begin{eqnarray}
m_{\sigma}^2 \,\bar{\sigma} &=& \sum_{B} g_{\sigma B} \, n_B^s - \frac{\partial U(\bar{\sigma})}{\partial \bar{\sigma}}, \\
m_{\sigma^*}^2 \,\bar{\sigma}^* &=&  \sum_{B} g_{\sigma^* B} \, n_B^s, \\ 
m_{\omega}^2 \,\bar{\omega} &=&  \sum_{B} g_{\omega B} \, n_{B}, \\
m_{\rho}^2 \,\bar{\rho} &=&  \sum_{B} g_{\rho B}(n_b)\,I_{3B} \, n_{B}, \\ 
m_{\phi}^2 \,\bar{\phi} &=&  \sum_{B} g_{\phi B} \, n_{B},
\end{eqnarray}
where $I_{3B}$ is the third component of isospin.

The scalar ($n_B^s$) and particle number ($n_B$) densities are given by
\begin{eqnarray}
n_{B}^s &=&  \frac{g_B}{2\pi^2} \int^{p_{F_B}}_0 p^2\, dp \,\frac{m_B^*}{E_B^*(p)}, \\
n_{B}   &=&  \frac{g_B\, p_{F_B}^3}{ 6 \pi^2 },
\end{eqnarray}
with $E_B^*(p) = \sqrt{p^2 + m_B^{*2}}$ and $g_B$ the spin degeneracy factor ($g_N=g_H=2$ for spin-1/2 nucleons and hyperons, $g_\Delta=4$ for spin-3/2 $\Delta$-isobars). The effective baryon mass, $m_B^*$, is modified by the scalar fields:
\begin{equation}
 m_B^* = m_B - g_{\sigma B}\,\bar{\sigma} - g_{\sigma^* B}\,\bar{\sigma}^*.
\end{equation}

We consider cold ($T=0$), catalyzed stellar matter. The composition of the star is determined by the requirements of weak interaction equilibrium and local charge neutrality. Since all baryons have baryon number $b_B=1$, the $\beta$-equilibrium conditions read
\begin{equation}
\mu_B = \mu_n - q_B \mu_e, 
\qquad
\mu_\mu = \mu_e,
\end{equation}
and charge neutrality imposes
\begin{equation}
\sum_B q_B n_B - n_e - n_\mu = 0,
\end{equation}
where $q_B$ is the electric charge of baryon $B$. Assuming that neutrinos have mean free paths larger than the stellar radius, they leave the system and we set $\mu_{\nu}=0$.

In the SW4L model, the chemical potential $\mu_B$ includes a rearrangement term, $\widetilde{R}$, arising from the density dependence of the isovector coupling, which ensures thermodynamic consistency \citep{Hofmann:2001aot}:
\begin{equation}
\mu_B = \sqrt{p^2_{F_B}+m_B^{*2}}  + g_{\omega B}  \bar{\omega} 
      + g_{\rho B}(n_b)  \bar{\rho}  I_{3B}  + g_{\phi B}  \bar{\phi}  + \widetilde{R},
\end{equation}
with
\begin{equation}
\widetilde{R} = \sum_B 
\left[\frac{\partial g_{\rho B}(n_b)}{\partial n_b}\right] 
I_{3B} \, n_B \,\bar{\rho}.
\end{equation}

Finally, the total pressure ($P$) and energy density ($\varepsilon$) of the system are obtained from the energy-momentum tensor. The total pressure is
\begin{equation}
\begin{split}
P &= \sum_B \frac{g_B}{6\pi^2} \int^{p_{F_B}}_0 
\frac{p^4\, dp}{\sqrt{p^2+m_B^{*2}}} \\
  &\quad + P_{\text{mesons}} + P_{\text{leptons}} + n_b \,\widetilde{R},
\end{split}
\label{eq:pressure_tot}
\end{equation}
where the mesonic and leptonic contributions are defined as
\begin{align}
P_{\text{mesons}} &= -\frac{1}{2} m_{\sigma}^2 \bar{\sigma}^2 - U(\bar{\sigma}) 
                    - \frac{1}{2} m_{\sigma^*}^2 \bar{\sigma}^{* 2} \nonumber \\
&\quad + \frac{1}{2} m_{\omega}^2 \bar{\omega}^2 
       + \frac{1}{2} m_{\rho}^2 \bar{\rho}^2
       + \frac{1}{2} m_{\phi}^2 \bar{\phi}^2, \\
P_{\text{leptons}} &= \sum_{L=e,\mu} \frac{1}{3\pi^2} 
\int^{p_{F_L}}_0 \frac{p^4\, dp}{\sqrt{p^2+m_L^2}}.
\end{align}
The total energy density is given by
\begin{equation}
\varepsilon = \sum_{i=B, L} \mu_i n_i - P.
\label{eq:EoS_final}
\end{equation}

\begin{table*}[t]
\caption{Global properties of neutron stars obtained for the EoS models defined in Table~\ref{tab:eos_params}. The columns list the maximum gravitational mass ($M_{\text{max}}$) and its corresponding radius ($R_{\text{max}}$). Additionally, we provide the radius ($R_{1.4}$), dimensionless tidal deformability ($\Lambda_{1.4}$), and  $g$-mode frequency ($\nu^g_{1.4}$) for a canonical $1.4~M_\odot$ star, relevant for comparison with GW170817 constraints.}
\label{tab:global_props}
\begin{ruledtabular}
\begin{tabular}{cccccc}
 & \multicolumn{2}{c}{Maximum Mass Configuration} & \multicolumn{3}{c}{Properties at $1.4~M_\odot$} \\
\cline{2-3} \cline{4-6}
EoS & $M_{\text{max}}$ ($M_\odot$) & $R_{\text{max}}$ (km) & $R_{1.4}$ (km) & $\Lambda_{1.4}$ & $\nu^g_{1.4}$ (Hz) \\
\colrule
eos-1 & 2.23 & 11.02 & 11.32 & 246.9 & 847.04 \\
eos-2 & 2.21 & 11.09 & 11.71 & 483.2 & 713.11 \\
eos-3 & 2.16 & 10.39 & 10.97 & 190.97 & 912.02 \\
eos-4 & 2.17 & 10.68 & 11.53 & 238.42 & 547.45 \\
\end{tabular}
\end{ruledtabular}
\end{table*}


\begin{figure*}[tb]
\centering
\includegraphics[width=\textwidth]{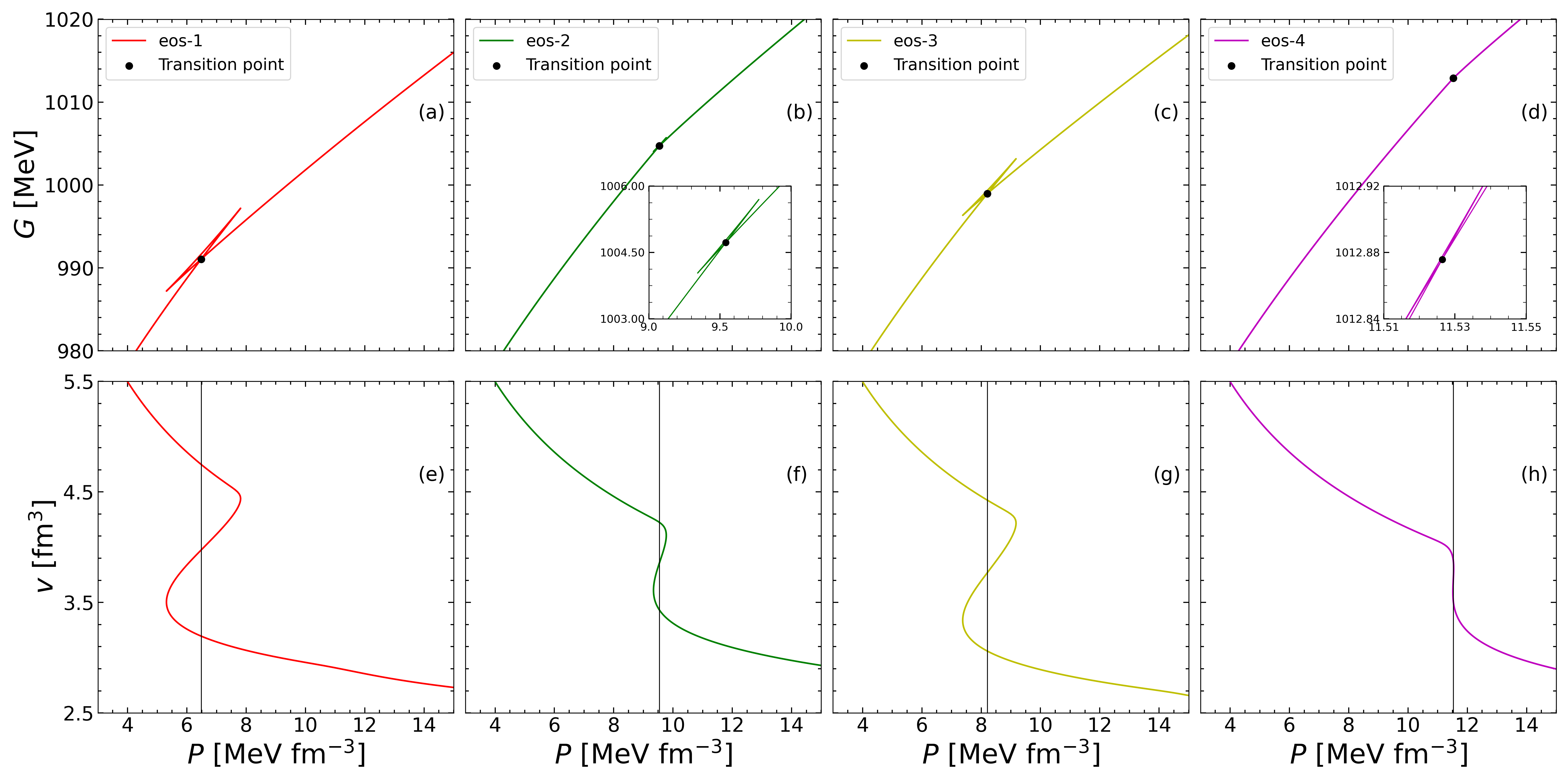}
\caption{(a)–(d) Gibbs free energy per baryon $G$ as a function of the pressure
$P$ for the four coupling sets of Table~\ref{tab:eos_params}. The multivalued
structure of $G(P)$ reflects the presence of two competing equilibrium
branches (a $\Delta$–free and a $\Delta$–rich phase); the black dot marks the
pressure $P_{\mathrm{tr}}$ at which their Gibbs free energies coincide,
signaling a hadron–hadron FOPT. (e)–(h) Volume per
baryon $v = n_b^{-1}$ as a function of $P$ for the same models. The
characteristic ``S-shaped'' loops indicate the spinodal region where
$dv/dP>0$ and homogeneous matter is mechanically unstable. The vertical solid
line denotes the transition pressure $P_{\mathrm{tr}}$ selected by the Maxwell
construction, at which the system jumps discontinuously from the low-density
to the high-density branch.}
\label{fig:G-P}
\end{figure*}

\section{Results} 
\label{sec:results}

In this section, we present the main thermodynamic and astrophysical consequences of varying the $\Delta$--meson coupling constants for the four representative EoS listed in Table~\ref{tab:eos_params}. We have explored several values of the poorly constrained ratios $x_{\sigma\Delta}$ and $x_{\omega\Delta}$, while keeping all other parameters of the SW4L parametrization fixed. The chosen sets extend slightly beyond the ranges discussed in 
Refs.~\cite{Malfatti:2020dba,Celi:2024doh}, 
but preserve coupling differences similar to those suggested in 
Ref.~\cite{Sedrakian:2022dra}.

\subsection{Thermodynamics and Microphysics of the $\Delta$–Induced Phase Transition}
\label{sec:results_A}

Consistent with the thermodynamic instabilities reported in Refs.~\cite{DeOliveira:2007eod,Deoliveira:2016pti,Lavagno:2019tii,Raduta:2021dan} using different hadronic models, we find that the SW4L parametrization also exhibits an intrahadronic FOPT between a $\Delta$-free hadronic phase and a $\Delta$-rich phase with an abundant $\Delta^{-}$ component when the couplings satisfy
\begin{equation}
 0.15 \lesssim x_{\sigma\Delta}-x_{\omega\Delta} \lesssim 0.2
 \quad \text{and} \quad
 x_{\sigma\Delta} \gtrsim 1.3.
\end{equation}
The coupling-difference window agrees closely with the range identified by Raduta~\cite{Raduta:2021dan} in the DDME2 framework and with the reduced vector-coupling regime of Refs.~\cite{DeOliveira:2007eod,Deoliveira:2016pti}, suggesting that the phenomenon is a robust feature of RMF models rather than an artifact of a particular parametrization. 
Within this interval, the thermodynamic observables display the canonical signatures of a first--order transition, including a discontinuity in the specific volume and the crossing of distinct branches of the Gibbs free energy per baryon. The strength of the discontinuity depends on the coupling difference: $\Delta x \equiv x_{\sigma\Delta}-x_{\omega\Delta}\sim 0.2$ yields a more pronounced energy--density jump, $\Delta\varepsilon \simeq 100$~MeV~fm$^{-3}$, whereas for smaller differences, $\Delta x\simeq 0.15$, the jump is weaker (see Table~\ref{tab:eos_params}). When the ordering of the couplings is inverted, $x_{\sigma\Delta} \lesssim x_{\omega\Delta}$, our scan indicates that no first--order transition is realized: $\Delta$ resonances may still appear, but only through a smooth, continuous onset.

This behavior admits a transparent thermodynamic interpretation. At zero temperature, under charge neutrality and $\beta$--equilibrium, the EoS can be parameterized by a single thermodynamic variable, for example the pressure $P$, but the underlying mean--field equations may nevertheless admit multiple microscopic solutions at the same $P$. These distinct solutions differ in their composition (e.g., $\Delta$--free versus $\Delta$--rich matter) and define separate thermodynamic branches of the Gibbs free energy per baryon, $G_\alpha(P) = \mu_{B,\alpha}(P)$, where the index $\alpha=1,2$ labels the $\Delta$--free and $\Delta$--rich branches, respectively. A first--order phase transition associated with the onset of $\Delta^{-}$ resonances then occurs at a pressure $P_{\rm tr}$ where the corresponding $G_\alpha(P)$ curves intersect, i.e., where the Gibbs free energies of the two phases coincide, $G_1(P_{\rm tr}) = G_2(P_{\rm tr})$; for $P < P_{\rm tr}$ the $\Delta$--free solution has the lower Gibbs free energy, while for $P > P_{\rm tr}$ the $\Delta^{-}$--rich solution is favored (see Figs. \ref{fig:G-P}(a)--(d)). In this description, the phase structure is entirely encoded in the competition between a small number of equilibrium branches, and the thermodynamically realized configuration at each pressure is simply the one with the lowest $G$.

Formally, this picture can be recast in terms of a Landau--type functional $G(P,\chi)$, where $\chi$ plays the role of an order parameter that quantifies the $\Delta^{-}$ content of the system, for instance the particle fraction $Y_{\Delta^{-}} = n_{\Delta^{-}}/n_b$. At fixed pressure, the $\Delta$-free and $\Delta$-rich branches correspond to distinct local minima of $G(P,\chi)$ with respect to $\chi$, i.e. configurations in which all particle populations satisfy the full set of chemical–equilibrium conditions. Values of $\chi$ away from these minima represent homogeneous states in which the $\Delta^{-}$ population is not in complete chemical equilibrium. Among the local minima, the one with the lowest Gibbs free energy identifies the stable phase, whereas any secondary minimum with higher $G$ corresponds to a metastable phase. In addition, a local maximum of $G(P,\chi)$  appears between the two minima; this maximum is associated with the third, intermediate branch in Figs. \ref{fig:G-P}(a)--(d) and represents a mechanically unstable configuration, as discussed below. Thus, the $\Delta$-free and $\Delta$-rich phases are mapped onto separate minima of the $G(P,\chi)$ functional, while the unstable branch is associated with the barrier separating them.
To understand why the functional $G(P,\chi)$ develops two minima for certain coupling choices, it is necessary to examine the microscopic energetics of the $\Delta^-$ threshold.

The microscopic origin of these two regimes can be understood by inspecting the effective single–particle energies. For the $\Delta^-$, the effective mass and chemical potential are given by the standard RMF expressions,
\begin{align}
 m_{\Delta^-}^* &= m_{\Delta^-} - g_{\sigma \Delta}\,\bar{\sigma}
                 - g_{\sigma^* \Delta}\,\bar{\sigma}^*,  \\
\mu_{\Delta^-} &= \sqrt{p^2_{F_{\Delta^-}}+m_{\Delta^-}^{*2}} + V_{\Delta^-},
\end{align}
where $V_{\Delta^-}$ denotes the sum of the vector contributions ($\omega,\rho,\phi$ and rearrangement terms). At zero temperature, the $\Delta^-$ Fermi sea starts to populate when the quasi–particle energy in the ground state equals the appropriate $\beta$–equilibrium combination of neutron and lepton chemical potentials,
\begin{equation}
\mu_{\Delta^-} = \mu_n + \mu_e 
  \simeq m_{\Delta^-}^* + V_{\Delta^-}.
\label{eq:delta_threshold}
\end{equation}
Below this threshold, $p_{F_{\Delta^-}}=0$ and the $\Delta^-$ Fermi sea is empty; above it, $p_{F_{\Delta^-}}^2 \propto (\mu_{\Delta^-}-V_{\Delta^-})^2 - m_{\Delta^-}^{*2}$ becomes positive and a nonzero fraction of $\Delta^-$ can appear by converting neutrons (and adjusting the charged components) through weak processes such as $n+e^- \rightarrow \Delta^- + \nu_e$.

Although our coupling choices do not lie in this regime, it is worth analyzing the behavior expected when the scalar attraction and vector repulsion are nearly balanced
($x_{\sigma\Delta}\!\approx\!x_{\omega\Delta}$). In that case, the combination $m_{\Delta^-}^*+V_{\Delta^-}$ varies smoothly with density: the decrease of $m_{\Delta^-}^*$ driven by the scalar field $\bar\sigma$ is largely compensated by the increase of the vector potential $V_{\Delta^-}$ driven by $\bar\omega$. In this case, crossing the threshold given by Eq.~\eqref{eq:delta_threshold} first occurs at a single density where $p_{F_{\Delta^-}}$ is very small. Adding a few $\Delta^-$ only mildly perturbs the mean fields and slightly lowers the Gibbs free energy relative to the purely nucleonic configuration. The minimum of $G(P,\chi)$ therefore shifts continuously from $\chi\simeq 0$ to small positive values: the system gradually converts neutrons into $\Delta^-$, the $\Delta$ Fermi seas fill up smoothly, and the EoS exhibits a continuous, crossover-like behavior with a single thermodynamic phase at each pressure.
This smooth-onset regime is well documented in the literature: representative calculations within different RMF variants confirm that the $\Delta$ population turns on continuously, with the principal effect being a gradual softening of the EoS and a reduction of $M_{\max}$ by up to $\sim 0.5\,M_\odot$, without any thermodynamic instability~\cite{Drago:2014ita,Cai:2015cda,Zhu:2016dri,Li:2018cbd,Ribes:2019ibd,Sedrakian:2023hbi}.

In contrast, when the scalar attraction dominates over the vector repulsion---i.e., in the coupling window identified above---the reduction of $m_{\Delta^-}^*$ with increasing density is much steeper, while the growth of $V_{\Delta^-}$ is comparatively weaker. In this regime, once a small population of $\Delta^-$ is present, its contribution to the scalar source term increases the magnitude of the scalar field $\bar\sigma$, further reducing $m_{\Delta^-}^*$ and lowering the cost of adding additional $\Delta^-$.
The physical origin of this self-reinforcement lies in the scalar field equation: the source of $\bar\sigma$ is the total scalar density, $\sum_i g_{\sigma i}\,n_{s,i}$, so that the appearance of a $\Delta^-$ population adds a term $g_{\sigma\Delta}\,n_{s,\Delta}$ that is amplified by the large value of $x_{\sigma\Delta}$.  The resulting increase in $\bar\sigma$ further lowers all baryon effective masses, but most strongly $m_{\Delta^-}^*$, closing the feedback loop. This scalar-driven self-reinforcement makes it energetically favorable, from the point of view of the scalar sector alone, to transfer a sizable fraction of baryon number from the neutron Fermi sea to the $\Delta^-$ (and other $\Delta$'s) Fermi seas. However, this gain is partly offset by additional costs in the vector and isovector sectors and in the leptonic  contributions:  the $\Delta$-rich configuration appears as a locally stable minimum of the Gibbs free energy in a finite pressure interval below $P_{\rm tr}$, without yet becoming the global minimum. 

As the pressure is increased towards $P_{\rm tr}$ from below, $\mu_n$ and $\mu_e$ continue to grow, and it becomes progressively more costly, in terms of single-particle energies, to accommodate additional baryon number in the neutron and lepton Fermi seas, whereas in the $\Delta$-rich phase, the strong scalar attraction further reduces $m_{\Delta^-}^*$ and allows part of this baryon number and charge to be carried by $\Delta^-$ states at a lower quasi-particle energy.
The scalar-driven gain in the \mbox{$\Delta$-rich} phase thus gradually compensates for the penalties from the vector, isovector, and rearrangement contributions; as the two minima approach degeneracy, it becomes energetically favorable to transfer a macroscopic fraction of baryon number from the neutron Fermi sea into the $\Delta^-$ Fermi sea. the Gibbs-energy difference between the two configurations shrinks to zero at $P_{\rm tr}$, and the FOPT takes place.

It is instructive to compare this microscopic interpretation with the diagnostic approaches used in earlier studies of $\Delta$-driven instabilities. Raduta~\cite{Raduta:2021dan} identified the onset of the FOPT by monitoring the sign change of the smallest eigenvalue of the free-energy curvature matrix. Lavagno and Pigato~\cite{Lavagno:2019tii} employed a related criterion based on the mechanical and chemical-diffusive susceptibilities of the multi-component system, while de~Oliveira \textit{et~al.}~\cite{DeOliveira:2007eod} observed the instability directly through the appearance of a negative compressibility, $dP/dn_b<0$.  All three approaches correctly \emph{diagnose} the presence of an instability but do not trace its physical origin. The Landau-type picture developed above complements these diagnostics by providing a transparent physical interpretation---the self-amplifying scalar feedback described in the preceding paragraphs---and by identifying the $\Delta^-$ particle fraction as the order parameter governing the transition. This understanding clarifies why the transition is controlled by the coupling \emph{difference} $\Delta x \equiv x_{\sigma\Delta}-x_{\omega\Delta}$ rather than by the individual coupling strengths: it is the imbalance between the self-amplifying scalar channel and the stabilizing vector repulsion that determines whether the feedback drives the system to a second minimum of $G(P,\chi)$ or is damped into a smooth crossover.


\begin{figure*}[tbh]
\centering
\includegraphics[width=\textwidth]{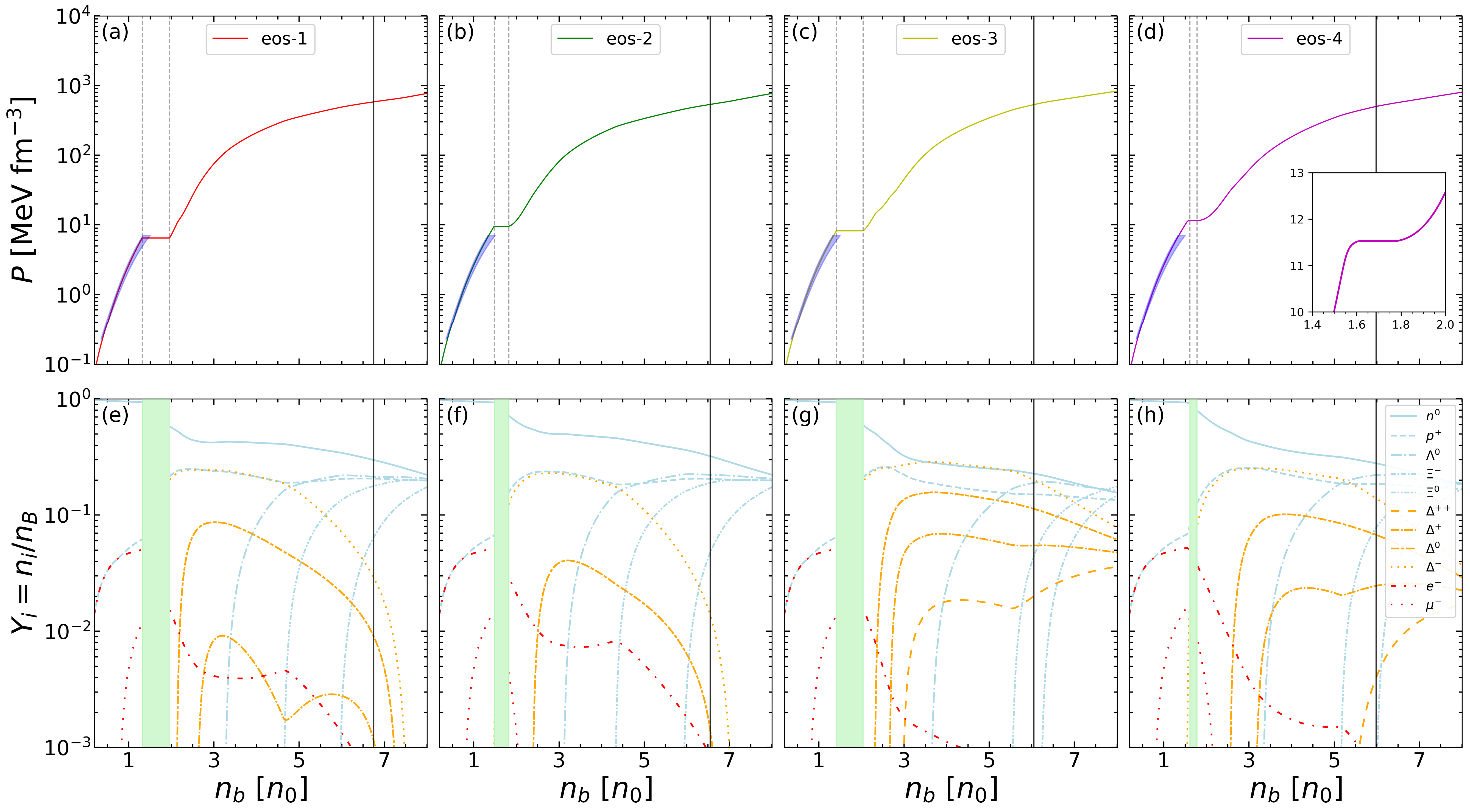}
\caption{(a)–(d) Total pressure as a function of the baryon number density $n_b$
for the four parameter sets defined in Table~\ref{tab:eos_params}. 
The horizontal plateaus correspond to the constant transition pressure 
$P_{\mathrm{tr}}$ obtained from the Maxwell construction; the endpoints of each 
plateau mark the baryon densities on the two sides of the discontinuity. 
The light-blue band indicates the region consistent with chiral effective field 
theory (cEFT) calculations at low densities based on Ref.~\cite{Drischler:2021lma}. 
(e)–(h) Corresponding particle fractions $Y_i = n_i/n_b$. 
The shaded vertical bands indicate the density gaps spanned by the pressure 
plateaus, i.e., the range of $n_b$ that is not realized in a homogeneous phase 
during the FOPT. The onset of the transition is associated with 
the rapid appearance of $\Delta^-$-isobars, which is accompanied by a sharp 
depletion of the electron and muon populations and a noticeable increase in the 
proton fraction, signaling a global reorganization of the Fermi seas across the 
transition. The black vertical line in each lower panel marks the central baryon 
density of the corresponding maximum-mass stellar configuration.}
\label{fig:EoS-PoP}
\end{figure*}

The consequences of this multivalued structure of the Gibbs free energy are explicitly realized in the behavior of the volume per baryon, $v=1/n_b$, displayed in Figs.~\ref{fig:G-P}(e)--(h). Since $v$ is thermodynamically conjugate to the pressure, $v = (\partial G / \partial P)_T$, the intersection of the two stable branches in the upper panels corresponds to a discontinuity in the baryon density at $P_{\mathrm{tr}}$, a hallmark of a FOPT.

The curves in Figs.~\ref{fig:G-P}(e)--(h) exhibit the characteristic ``S-shaped'' or van der Waals--like loops typical of mean-field descriptions of FOPTs. The \mbox{$\Delta$--free} and $\Delta$--rich branches, which correspond to the local minima of $G(P,\chi)$, act as the stable or metastable segments of the EoS, where the mechanical stability condition $dP/dn_b>0$ (or equivalently $dv/dP<0$) is satisfied. Conversely, the intermediate branch connecting these two regimes corresponds to the local maximum of the Gibbs functional discussed above. In the $v$–$P$ plane, this segment exhibits a positive slope, $dv/dP>0$, indicating a region of mechanical instability (spinodal region) where small fluctuations would grow and drive the system towards phase separation.

Physically, the system follows the Maxwell construction: it evolves along the $\Delta$--free curve up to $P_{\mathrm{tr}}$, then jumps discontinuously across the spinodal region to the $\Delta$--rich curve, ensuring that it always resides in the global minimum of $G$. This thermodynamic trajectory is illustrated in Figs.~\ref{fig:EoS-PoP}(a)--(d), where the pressure is plotted as a function of the baryon number density $n_b$. The unstable and metastable segments of the ``S-shaped'' loops are replaced by horizontal plateaus of constant pressure $P_{\mathrm{tr}}$, which connect the low-density hadronic phase to the high-density $\Delta$-rich phase. The width of these plateaus represents the density jump $\Delta n_b$ across the phase transition, and coincides with the shaded vertical bands in Figs.~\ref{fig:EoS-PoP}(e)--(h).

The microscopic drivers of this discontinuity are revealed by inspecting the particle fractions $Y_i = n_i/n_b$ shown in Figs.~\ref{fig:EoS-PoP}(e)--(h). In all cases where a FOPT occurs, the phase boundary is tightly linked to the sudden emergence of $\Delta^-$-isobars. To the left of the density-gap band, the composition is that of standard neutron-star matter: neutrons, a smaller proton fraction, and electrons and muons in $\beta$-equilibrium. As the density increases, the neutron and electron chemical potentials grow and eventually satisfy the threshold condition $\mu_{\Delta^-} = \mu_n + \mu_e$, allowing the $\Delta^-$ Fermi sea to start forming.

The transition, however, is not merely the appearance of a new species, but a drastic reorganization of the Fermi seas. The system lowers its Gibbs free energy by converting energetic neutrons and leptons into $\Delta^-$ baryons that occupy states near the bottom of their effective mass shell. Immediately to the right of the green band, the $\Delta^-$ fraction rises to values of order $10^{-1}$, accompanied by a sharp drop in the neutron fraction. The electron and muon populations are strongly suppressed as well: once negatively charged baryons are available, charge neutrality can be maintained with far fewer high-$\mu_e$ and high-$\mu_\mu$ leptons. At the same time, the proton fraction increases noticeably across the transition. This behavior reflects the tendency of the strong interaction to reduce isospin asymmetry in the baryonic sector: a configuration with a comparably large population of $p^+$ and $\Delta^-$, plus only a small lepton component, can satisfy both charge neutrality and $\beta$-equilibrium while keeping the baryons closer to isospin symmetry and lowering the overall energy per particle. As a result, in the $\Delta$-rich phase the abundances of $\Delta^-$ and protons remain remarkably close over a broad density interval, a direct consequence of charge-neutrality and $\beta$-equilibrium constraints operating in a regime where leptons play only a subdominant role.
This abrupt reorganization of the Fermi seas is a direct macroscopic manifestation of the scalar feedback mechanism discussed above: once the two Gibbs-energy minima become degenerate at $P_{\rm tr}$, the transfer of baryon number into the $\Delta^-$ Fermi sea is energetically self-reinforcing and proceeds as a discontinuous jump rather than a gradual crossover.

Finally, the variation of the density-gap width from Fig.~\ref{fig:EoS-PoP}(e) to \ref{fig:EoS-PoP}(h) closely mirrors the coupling dependence discussed above. For the parameter sets with the largest scalar–vector difference [Figs.~\ref{fig:EoS-PoP}(e) and \ref{fig:EoS-PoP}(g)], the strong scalar attraction leads to a broad forbidden region and a large density jump. As the difference between scalar and vector couplings decreases [Figs.~\ref{fig:EoS-PoP}(f) and \ref{fig:EoS-PoP}(h)], the energy gain from introducing $\Delta^-$ diminishes, the pressure plateau narrows, and the density gap shrinks. In the weakest-transition case [Fig.~\ref{fig:EoS-PoP}(h)], the green band and the associated wiggle in the pressure curve are already very narrow: the $\Delta^-$ fraction remains essentially zero on the $\Delta$-free side and rises steeply only as the thin density-gap is crossed. Any tiny pre-transitional $\Delta^-$ fraction visible at the edge of the band is within our numerical resolution of the plateau and does not represent an additional phase. This case therefore represents a very weak but still finite FOPT, lying close to the regime where the phase change would become a smooth crossover.

\begin{figure*}[tbh]
\centering   
\includegraphics[width=\textwidth]{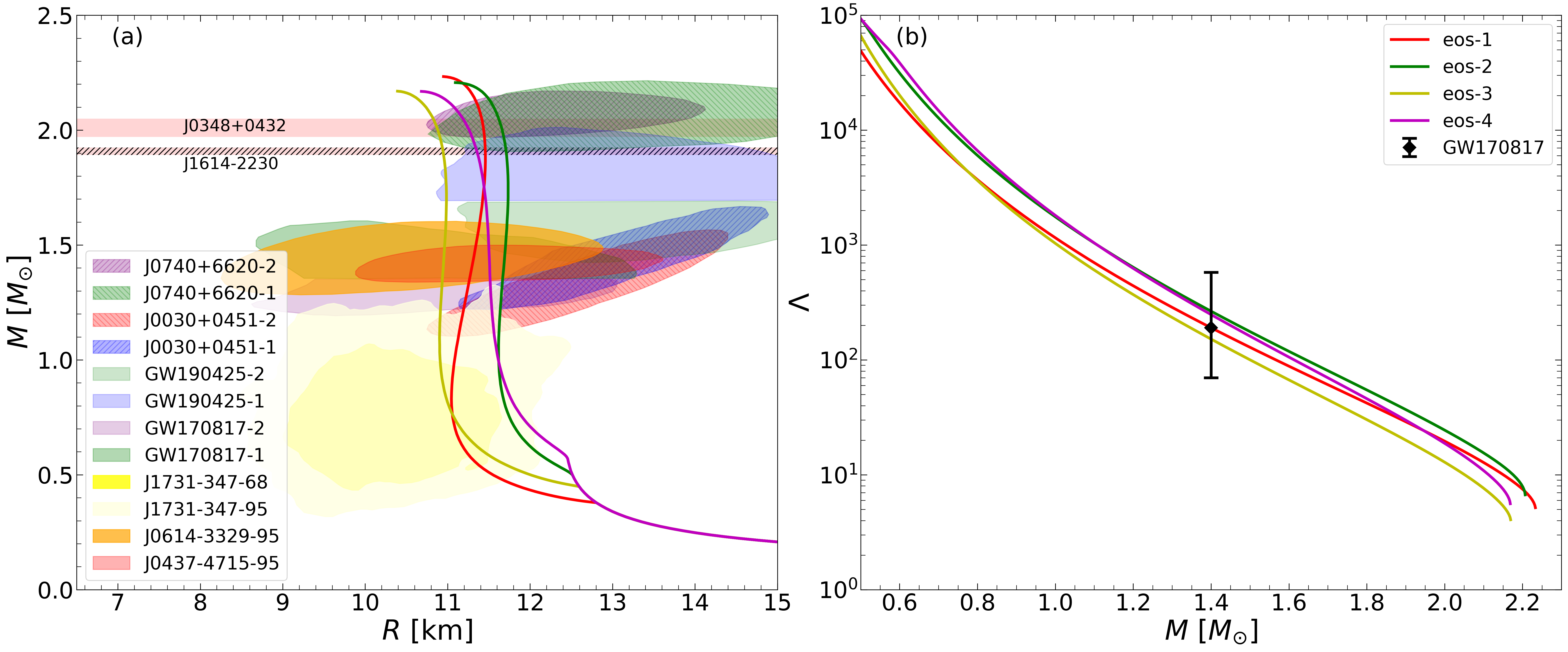}
\caption{Macroscopic structure of neutron stars for the selected EoS parametrizations. (a) Mass–radius relations compared with observational constraints from massive pulsars (shaded horizontal bands) and from NICER and GW170817 (colored confidence regions). The change in slope (``knee'') of the curves reflects the onset of the $\Delta$-rich phase in the stellar core; the sharpness of this feature correlates with the strength of the FOPT. (b) Dimensionless tidal deformability $\Lambda$ as a function of stellar mass. The black diamond with error bars indicates the constraint on $\Lambda_{1.4}$ from the binary neutron star merger GW170817. All four EoS yield compact stars compatible with current multimessenger observations~\citep{Abbott:2018gmo}}  
\label{fig:MR}
\end{figure*}

\subsection[Macroscopic Signatures: $M$, $R$, $\Lambda$ and $g$-modes]{Macroscopic Signatures: Mass--Radius Relations, Tidal Deformability and $g$-modes}
\label{sec:results_B}

The thermodynamic properties and phase structure discussed above have direct consequences for the macroscopic observables of neutron stars. By solving the Tolman–Oppenheimer–Volkoff (TOV) equations together with the tidal deformability equations, we obtain the mass–radius ($M$–$R$) relations and the dimensionless tidal deformability $\Lambda$ shown in Fig.~\ref{fig:MR}.

Figure~\ref{fig:MR}(a) displays the $M$–$R$ curves for the four representative EoS. At low central densities \mbox{($M \lesssim 0.5\,M_\odot$)}, all models lie on a common branch that describes purely nucleonic stars with relatively large radii. Once the central density exceeds the transition range identified in Sec.~\ref{sec:results_A} and a $\Delta^-$–rich core forms, the $\Delta$ contribution softens the EoS and alters the star’s response to gravity. This softening manifests itself as a characteristic change in slope (a ``knee'') in the $M$–$R$ curve, signaling the onset of a $\Delta$-rich core. For masses above the knee, the configurations are purely hadronic \emph{hybrid} stars: they contain an inner core dominated by $\Delta^-$ resonances (and, at higher densities, by other $\Delta$ states and hyperons), separated by a density discontinuity from an outer core of nucleonic matter.

The morphology of this knee closely reflects the order and strength of the phase transition. For the parameter sets with the largest density gap (eos-1 and eos-3), the sizeable energy-density discontinuity \mbox{($\Delta\varepsilon \simeq 100$~MeV~fm$^{-3}$)} produces a pronounced reduction in radius over a narrow mass interval: as the stellar core converts to the more compact $\Delta$-rich phase, gravity drives a substantial contraction of the star at nearly fixed mass. For smaller coupling differences (eos-2 and eos-4) the transition is weaker and the kink is correspondingly milder. 

Despite the early onset of $\Delta^-$-isobars and the associated softening at intermediate densities, all four EoS remain compatible with current mass measurements of heavy pulsars. The sequences reach maximum masses in the range $M_{\rm max} \approx 2.15$–$2.25\,M_\odot$, satisfying the constraints from PSR J1614-2230  and PSR J0348+0432. This shows that the vector repulsion in the high-density $\Delta$-rich phase is strong enough to support massive stars against collapse, even though the radii of intermediate-mass stars are reduced by the transition.

The predicted radii are also consistent with multimessenger constraints. For all four EoS, the onset density of the $\Delta^-$ resonance is such that canonical $1.4\,M_\odot$ stars lie on the $\Delta$-rich branch. As a result, the models yield relatively compact radii $R_{1.4}\approx 11$–$12$~km, compatible with NICER measurements of PSR J0030+0451 and PSR J0740+6620 and with the posterior regions inferred from GW170817. Moreover, the presence of the ``knee'' in the $M$–$R$ relation allows all four models to accommodate the very low mass inferred for the compact object in J1731–347. The early $\Delta$ onset thus provides a natural microphysical mechanism to reconcile large maximum masses with comparatively small radii at $M\sim 1.4\,M_\odot$.

The morphology of the $M$--$R$ curves obtained here is qualitatively
consistent with the results of Raduta~\cite{Raduta:2021dan}, who
reported a similar knee structure and comparable maximum masses
($M_{\max} \approx 2.0$--$2.3\,M_\odot$) in the DDME2 model for the
same coupling-difference window.  Our predicted radii
($R_{1.4}\approx 11$--$12$~km) are also in line with the radius
reduction reported in Ref.~\cite{Raduta:2021dan}.  The agreement between
two distinct RMF parametrizations---density-dependent couplings (DDME2)
and the SW4L model employed here---reinforces the conclusion that the
$\Delta$-induced phase transition is a robust phenomenon across
different RMF frameworks.
De~Oliveira \textit{et~al.}~\cite{DeOliveira:2007eod} also obtained a
``kink'' in $M(R)$ within a nonlinear Walecka model, although their
analysis predated the multimessenger constraints now available.

Figure~\ref{fig:MR}(b) illustrates the dimensionless tidal deformability, $\Lambda$, as a function of stellar gravitational mass. All four EoS demonstrate excellent agreement with the limits derived from the binary neutron star merger GW170817, falling well within the observational confidence intervals. A clear hierarchy emerges among the models: those featuring the strongest FOPTs and widest density gaps (eos-1 and eos-3) systematically predict the lowest values of $\Lambda$. This behavior is a direct consequence of the extreme sensitivity of tidal deformations to stellar compactness, which scales as $\Lambda \propto k_2 (R/M)^5$. The enhanced softening and radius reduction induced by the sharp onset of the $\Delta$-rich phase suppress the quadrupolar response of the star to the external tidal field, thereby yielding lower $\Lambda$ values for the models with more pronounced density discontinuities.

Raduta~\cite{Raduta:2021dan} reported that the $\Delta$-induced transition produces a reduction of order a few percent in $\Lambda$ relative to the corresponding $\Delta$-free EoS. Our results are consistent with this finding, although the absolute values of $\Lambda_{1.4}$ differ between models (see Table~\ref{tab:global_props}) due to the different saturation properties and high-density stiffness of SW4L compared to DDME2.  For context, calculations of $\Delta$-admixed stars with a \emph{smooth} onset of $\Delta$ particles~\cite{Li:2018cbd,Ribes:2019ibd} yield values of $\Lambda_{1.4}$ higher than those satisfying the constraint imposed by GW170817. This indicates that the FOPT provides an additional, non-perturbative mechanism for compactifying the star beyond the continuous softening already induced by the $\Delta$ population, thus producing small tidal deformabilities.

Having established the signatures of the $\Delta$-driven phase transition on the static stellar structure, we now turn to its imprint on the oscillation spectrum.  We first assess the dynamical stability of the equilibrium configurations by computing the eigenfrequencies of the fundamental radial oscillation mode. Our numerical results show that stability is lost precisely at the maximum-mass configuration, where the fundamental mode satisfies $\omega_0^2 = 0$. This behavior is found independently of the assumed microphysics of the phase-conversion process at the interface: the stability boundary remains located at the peak of the $M$–$R$ curve in both the rapid- and slow-conversion limits~\cite{Pereira:2017rmp}. Moreover, the onset of the $\Delta$-rich core does not induce any additional instability: hybrid configurations remain dynamically stable immediately after the phase transition sets in. This is consistent with the Seidov criterion~\cite{Seidov:1971tso}, which we have verified explicitly for our EoS. In all four cases, the energy-density discontinuity satisfies the condition $\Delta\varepsilon < \Delta\varepsilon_{\rm crit} \equiv (\varepsilon_{\rm onset} + 3 P_{\rm tr})/2$, ensuring that the appearance of a macroscopic $\Delta$-rich core does not destabilize the star.

The density discontinuity that characterizes the $\Delta$-induced phase transition also gives rise to a composition gravity mode, or $g$-mode, whose restoring force is buoyancy at the phase boundary rather than pressure gradients.
We have computed the complex eigenfrequencies $\omega _g = 2\pi\nu_g + i/\tau_g$ of these non-radial ($\ell=2$) perturbations within the linearized General Relativity framework of Ref.~\cite{lindblom:1983tqo,detweiler:1985otn}.

\begin{figure}[tb]
\centering
\includegraphics[width=\columnwidth]{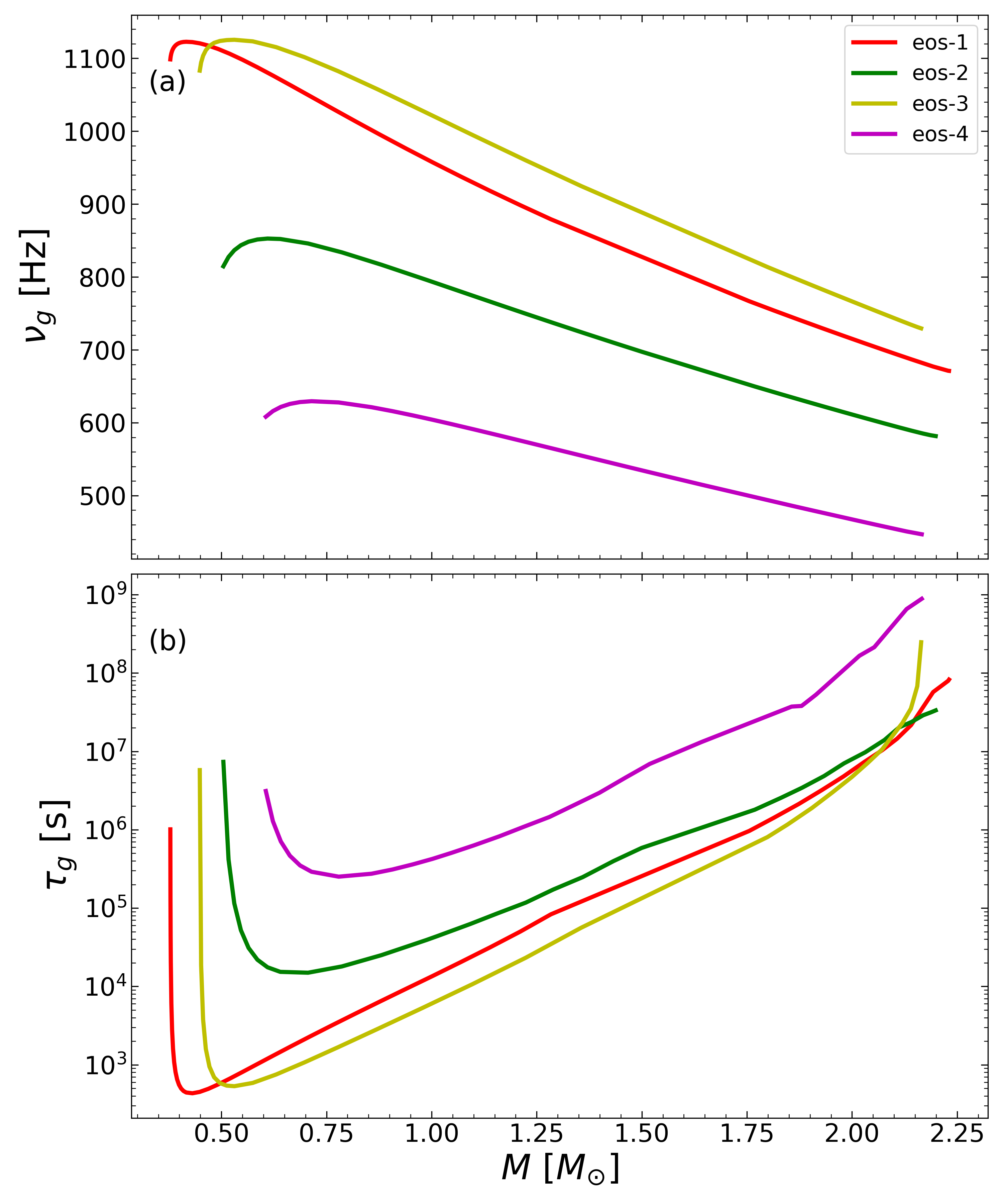}
\caption{Properties of the composition $g$-mode associated with the hadron–hadron phase-transition interface. (a) Gravitational wave frequency, $\nu_g$, as a function of stellar mass. (b) Damping time due to gravitational-wave emission, $\tau_g$, as a function of the stellar mass.}
 \label{fig:g-modes}
\end{figure}

Figure~\ref{fig:g-modes}(a) displays the oscillation frequency $\nu_g$ as a function of stellar mass. As expected for a discontinuity-driven mode, the
$g$-mode exists only in stars massive enough to develop a macroscopic $\Delta$-rich core; for the hybrid configurations, the frequencies lie in the typical range $\nu_g \sim 400$–$1100$~Hz. The behavior of the curves reveals a clear correlation between the microphysics of the transition and the mode frequency. Analytic estimates for $g$-modes localized at a sharp density jump
indicate that the squared frequency scales with the local gravitational acceleration and the density contrast at the interface, $\omega_g^2 \sim
\Delta\varepsilon/\varepsilon$~\cite{Sotani:2001ddo, Miniutti:2002nro, Rodriguez:2025cog}. This scaling is well
supported by our results: models with a sharper transition and larger energy–density jump (eos-1 and eos-3) exhibit the highest frequencies, reaching values above $1$~kHz for low-mass hybrid stars. Conversely, eos-4, which features a very narrow density gap approaching the crossover limit, systematically yields lower frequencies ($\nu_g \lesssim 600$~Hz). In all cases, the mild decrease of $\nu_g$ with increasing stellar mass reflects the outward migration of the phase boundary and the evolution of the local gravitational field at the interface. In simple Newtonian models of incompressible stars with a sharp transition at $r=r_{\rm tr}$, the quadrupolar $g$-mode frequency indeed scales not only with $\sqrt{\Delta\varepsilon/\varepsilon}$, but also with $\sqrt{1-(r_{\rm tr}/R)^5}$~\cite{counsell:2025imi}, illustrating explicitly the combined sensitivity to the density jump and to the location of the interface inside the star.

Figure~\ref{fig:g-modes}(b) shows the damping time, $\tau_g$, due to gravitational-wave emission. We obtain very long damping times, $\tau_g \sim 10^3$–$10^9$~s, several orders of magnitude larger than those of typical fluid pressure modes \mbox{($\tau_f \sim 0.1$–$1$~s)}. The damping times exhibit qualitatively similar behavior to those obtained in Ref. \cite{Mariani:2022omh} for hybrid stars with quark matter in their cores. These long timescales indicate that the $g$-modes are effectively trapped in the deep interior and couple only weakly to the spacetime metric. Since the eigenfunctions are localized near the phase boundary and involve relatively small mass rearrangements, they generate only a modest quadrupole moment. Consequently, the direct detection of the associated ringdown signal is unlikely with the current generation of detectors.

The theory of $g$-modes associated with density discontinuities in neutron stars was developed by Finn~\cite{Finn:1987gmi} and McDermott~\cite{McDermott:1990ddg}, who showed that each sharp interface supports a single mode per spherical harmonic order $\ell$.  The application to hadron--quark phase boundaries has since been pursued by several groups~\cite{Sotani:2001ddo, Miniutti:2002nro, Flores:2013dha, Tonetto:2020dgm, Rodriguez:2021hsw, Rodriguez:2025cog, Mariani:2022omh}. In all these studies, the density discontinuity was assumed to originate from a deconfinement transition to quark matter. Counsell \textit{et~al.}~\cite{counsell:2025imi} and Pereira \textit{et~al.}~\cite{pereira:2025dti} recently studied the tidal excitation of interface modes in a model-agnostic setting informed by chiral effective field theory, showing that the resonant excitation could induce a detectable phase shift in the gravitational waveform, potentially observable with LIGO~A+ and third-generation detectors such as Cosmic Explorer and the Einstein Telescope. Our results demonstrate that the same interface-mode phenomenology arises from a purely hadronic mechanism: the density discontinuity produced by $\Delta$-isobar formation provides a concrete microphysical source for the sharp interface that is a prerequisite for the scenarios analyzed in Refs.~\cite{counsell:2025imi,pereira:2025dti}. The fact that our $\nu_g$ values ($\sim 400$--$1100$~Hz) overlap with the range $\sim 500$--$1000$~Hz obtained for sharp hadron--quark interfaces in MIT bag model calculations~\cite{Tonetto:2020dgm} and with the low-pressure hadron--quark transitions presented in Ref. \cite{Ranea:2018omo,Rodriguez:2021hsw} and in ``family~IV'' of Ref.~\cite{Rodriguez:2025cog} is a direct manifestation of the masquerade at the level of the oscillation spectrum: the $g$-mode frequency is primarily controlled by the fractional density contrast $\Delta\varepsilon/\varepsilon$ and the radial location of the interface, rather than by the microscopic nature of the high-density phase.

\subsection{Comparison with Model-Agnostic Bayesian Constraints}
\label{sec:bayesian_comparison}

Having characterized the static and dynamical signatures of the
$\Delta$-driven phase transition, we now compare our four EoS models
with the constraints derived from recent model-agnostic Bayesian
analyses of the neutron-star EoS.  Before proceeding, we note that
different studies quantify the strength of a FOTP using different variables---fractional baryon-density jump
$\Delta n/n$, absolute density jump $\Delta n$ (in units of
$n_{0}$), energy-density discontinuity $\Delta\varepsilon$
(in MeV~fm$^{-3}$), or latent energy per baryon
$\Delta(E/N)$---so that cross-comparisons require care.

Gorda \textit{et~al.}~\cite{Gorda:2022lsk} constructed a large
ensemble of piecewise-polytropic EoSs with an explicit FOPT, anchored to chiral EFT at nuclear density and to
perturbative QCD at high densities, and subjected them to
astrophysical constraints from massive pulsars and gravitational-wave
observations.  They map the allowed parameter space of transition
onset density and latent energy, finding that transitions beginning
below ${\sim}\,2\,n_0$ can accommodate neutron-star configurations
with larger radii, while strong transitions at higher densities are
increasingly disfavoured; they also identify a small region of
parameter space allowing twin-star solutions that only marginally
survives current constraints.  Brandes \textit{et~al.}~\cite{Brandes:2023bob}
performed a Bayesian inference of the speed of sound incorporating
data from the heavy ($2.35\,M_\odot$) black-widow pulsar
PSR~J0952--0607 and, through a systematic Bayes-factor analysis,
found evidence \emph{against} low averaged sound speeds
($c_s^2 \leq 0.1$) throughout the density range realized in
neutron-star cores; within their 68\% posterior credible bands, only
a weak transition with $\Delta n/n \lesssim 0.2$ remains compatible
with the data.  Importantly, neither of these studies explicitly
models the FOPT itself~\cite{Komoltsev:2024lcr}.
Komoltsev~\cite{Komoltsev:2024lcr} performed a fully Bayesian
inference using Gaussian processes with explicit inclusion of FOPTs
and found that the current data cannot differentiate between a smooth
crossover and a FOPT: the Bayes factors between
the two scenarios are of order unity (see Table~II of
Ref.~\cite{Komoltsev:2024lcr}), and roughly $91\%$ of the total
Bayesian evidence consists of EoSs that exhibit some form of phase
change.  The posterior distribution in the
$n_{\rm PT}$--$\Delta n$ plane (Fig.~3 of
Ref.~\cite{Komoltsev:2024lcr}) reveals two distinct islands of
non-negligible posterior weight for FOPTs inside stable neutron
stars: an early-onset region at
$n_{\rm PT}\in [1,2]\,n_{0}$ with $\Delta n \lesssim
1.2\,n_{0}$, and a higher-density region at
$n_{\rm PT}\in [3,6]\,n_{0}$, separated by a gap imposed by
the mass constraints.

These results can be compared directly with our four EoS models.
The fractional baryon-density jumps span a broad range:
$\Delta n / n \approx 0.11$ (eos-4), $0.23$ (eos-2), $0.45$
(eos-3), and $0.48$ (eos-1); equivalently, the absolute density
jumps are $\Delta n \approx 0.17$, $0.34$, $0.63$, and
$0.64\,n_{0}$, with onset densities \mbox{$n_{\rm onset}\approx
(1.3$--$1.6)\,n_0$}.  All four models therefore fall squarely within
the early-onset island identified by
Komoltsev~\cite{Komoltsev:2024lcr}, and well below the
$\Delta n \lesssim 1.2\,n_{0}$ ceiling above which the FOPT
would destabilize the star.  The weakest-transition model, eos-4
($\Delta\varepsilon \approx 27$~MeV~fm$^{-3}$, $\Delta n/n \approx
0.11$), comfortably satisfies the Brandes \textit{et~al.}\ bound;
eos-2 ($\Delta\varepsilon \approx 55$~MeV~fm$^{-3}$, $\Delta n/n
\approx 0.23$) lies at the edge of the 68\% credible region.  For
the strongest transitions, eos-1 and eos-3
($\Delta\varepsilon \approx 100$~MeV~fm$^{-3}$, $\Delta n/n
\approx 0.45$--$0.48$), the density jumps exceed the $\Delta n / n
\sim 0.2$ bound reported in Ref.~\cite{Brandes:2023bob}.

The interpretation of this apparent tension requires care.
The Bayesian bounds of
Refs.~\cite{Gorda:2022lsk,Brandes:2023bob} are statements about the
\emph{prior-weighted volume} of a particular model-agnostic EoS
parametrization that passes the observational filters, not direct
statements about the physical likelihood of a given transition
strength.  Both analyses employ simplified representations of the
high-density EoS---piecewise polytropes~\cite{Gorda:2022lsk} or
segment-wise linear interpolations of
$c_s^2$~\cite{Brandes:2023bob}---whose parameters are varied over
broad, essentially uninformed priors.  In such a setup, a large
density jump requires a finely coordinated sequence of EoS
properties: the pre-transition segment must be stiff enough to
produce the observed radii, the transition must be followed by a
rapid stiffening that restores $M_{\max}>2\,M_\odot$, and the pQCD
constraint must be met at asymptotically high densities.  Within a
generic parametrization, relatively few randomly drawn parameter
combinations satisfy all these requirements simultaneously, leading
to a small posterior weight for strong transitions.  This is,
however, a property of the \emph{parametrization and the prior
measure}: it reflects the fact that the simplified model space does
not encode the microscopic correlations that a concrete physical
mechanism enforces among the eos
segments.
In the case of the $\Delta$-driven transition, these correlations are
manifest: the same scalar-field self-amplification that produces a
large $\Delta n/n$ also guarantees a rapid post-transition stiffening, because the vector repulsion among $\Delta$ isobars
grows in step with their population.  In a model-agnostic parametrization these two features are treated as independent parameters, so their correlation is invisible to the prior and appears only as a suppression of posterior volume.
That the current data are, in fact, unable to favor a smooth crossover
over a FOPT, as demonstrated by the order-unity Bayes factors of
Ref.~\cite{Komoltsev:2024lcr}, reinforces this conclusion.

Our microscopic calculation provides a concrete illustration: eos-1
and eos-3, whose density jumps lie outside the 68\% credible bands
of Ref.~\cite{Brandes:2023bob}, nevertheless satisfy every current
observational constraint---$M_{\max}>2\,M_\odot$, $R_{1.4}\approx
11$--$12$~km, $\Lambda_{1.4}\approx 190$--$250$, compatibility with
the GW170817 tidal-deformability limits, and consistency with chiral
EFT at low density---while lying within the allowed region of the
Komoltsev~\cite{Komoltsev:2024lcr} posterior.  Furthermore, in all
four cases the energy-density discontinuity remains below the Seidov
threshold, so that no disconnected twin-star branch appears,
consistent with the finding that twin-star solutions carry vanishing
Bayesian evidence~\cite{Komoltsev:2024lcr,Gorda:2022lsk}.  This
suggests that microphysics-informed EoS models can populate regions
of the functional EoS space that are undersampled by model-agnostic
priors, precisely because the physical correlations among transition
properties that ensure astrophysical viability are not encoded in
generic parametrizations.

\section{Summary and Discussion}
\label{sec:summary}

In this work, we have carried out a systematic study of $\Delta$-isobar degrees of freedom in high-density neutron-star matter within a relativistic mean-field framework. By exploring the parameter space of the scalar and vector meson–$\Delta$ couplings, we have identified a specific window of interaction strengths—characterized by a dominant scalar attraction ($x_{\sigma\Delta} \gtrsim 1.3$) and a moderate vector-repulsion difference ($0.15 \lesssim x_{\sigma\Delta}-x_{\omega\Delta} \lesssim 0.2$)—in which the onset of $\Delta$ formation manifests as a FOPT.
The $\Delta$-driven FOPT has now been identified in four independent RMF frameworks: a nonlinear Walecka model~\cite{DeOliveira:2007eod,Deoliveira:2016pti}, the SFHo model with a Gibbs construction~\cite{Lavagno:2019tii}, the DDME2 covariant density functional with a Maxwell construction~\cite{Raduta:2021dan}, and the SW4L parametrization employed in this work.  The fact that the instability arises for a similar coupling-difference window in all four cases ($\Delta x \approx 0.15$--$0.2$) suggests that it is an intrinsic feature of the RMF description of $\Delta$-admixed matter rather than an artifact of a particular parametrization, although a systematic study across a broader class of interactions (e.g., chiral EFT-based functionals) would be desirable.

The thermodynamics of this transition is driven by a non-linear feedback mechanism in the scalar sector: the appearance of $\Delta^-$ isobars enhances the scalar mean field, which in turn reduces the effective baryon mass, energetically favoring the further conversion of neutrons into $\Delta^-$ states. This produces a van der Waals–like instability loop in the EoS, resulting in a macroscopic density gap where homogeneous matter is mechanically unstable. Through a Maxwell construction, we showed that this reorganization of the Fermi seas leads to the formation of a $\Delta$-rich core separated from the outer nucleonic mantle by a sharp density discontinuity.

From an astrophysical perspective, the consequences of this intrahadronic phase transition are significant and help address several current tensions in neutron-star phenomenology. The early onset of the $\Delta^-$ resonance softens the EoS at intermediate densities, ensuring that canonical $1.4\,M_\odot$ stars remain compact ($R_{1.4} \approx 11$–$12$~km) and exhibit low tidal deformabilities ($\Lambda_{1.4} \approx 200$–$500$), in very good agreement with the multimessenger constraints from GW170817 and NICER data. At the same time, the stiffness of the EoS is restored at higher densities due to the vector repulsion among isobars, allowing the $M$-$R$ sequences to support maximum masses well above $2\,M_\odot$ and thus satisfy the existence of massive pulsars such as PSR J0348+0432. Furthermore, the characteristic ``knee'' or change in slope in the mass–radius relation induced by the phase transition provides a plausible microphysical explanation for very low-mass compact objects, such as the central compact object in HESS J1731–347, without invoking non-hadronic phases.

A notable outcome of our analysis is the close phenomenological similarity between the $\Delta$-induced phase transition and the deconfinement transition to quark matter. Our results demonstrate that purely hadronic models can reproduce the ``hybrid star'' phenomenology—compact radii, mass–radius kinks, and density discontinuities—traditionally associated with quark cores. This ``masquerade'' extends to the dynamical spectrum of the star. We have shown that the density discontinuity at the $\Delta$-formation threshold supports a gravity mode \mbox{($g$-mode)} with frequencies in the range \mbox{$\nu_g \sim 400$–$1100$~Hz}. These frequencies overlap substantially with those predicted for hadron–quark phase transitions, indicating that the presence of such a mode is primarily sensitive to the existence and strength of a density jump, rather than to the microscopic nature of the high-density phase itself.

The detectability of these modes presents both a challenge and an opportunity for future gravitational-wave astronomy. The calculated damping times for the
\mbox{$g$-modes} are extremely long ($\tau_g \gtrsim 10^3$~s), implying a very weak
coupling to the spacetime metric and making the direct detection of $g$-mode
ringdown immediately after a merger unlikely with current sensitivities.
However, the presence of a sharp discontinuity opens the possibility of
resonant excitation of these modes by tidal forces during the late inspiral
phase of a binary coalescence, which could induce a phase shift in the
gravitational waveform potentially detectable by next-generation
interferometers such as the Einstein Telescope or Cosmic Explorer.

While the frequency degeneracy complicates the distinction between $\Delta$-rich cores and quark cores via asteroseismology alone, the combination of static and dynamical observables may eventually break the degeneracy. For instance, if future observations were to constrain stellar radii at the percent level while simultaneously detecting a mode signature, the specific correlation between $R$ and $\nu_g$ predicted by each class of models could serve as a discriminator. Additionally, thermal evolution could provide complementary clues: direct Urca processes involving isobars (e.g., $\Delta^- \to n + e^- + \bar{\nu}_e$) differ from quark neutrino emissivity channels, potentially leading to distinct cooling tracks.

Our four EoS models are also broadly consistent with recent model-agnostic Bayesian constraints on FOPTs. All four fall within the early-onset island of the Komoltsev~\cite{Komoltsev:2024lcr} posterior; the weaker transitions (eos-2 and~4) satisfy the Brandes \textit{et~al.}~\cite{Brandes:2023bob} bound on $\Delta n/n$, while the stronger ones (eos-1 and~3) exceed it yet pass all current astrophysical constraints, illustrating that microphysics-informed correlations can populate EoS regions undersampled by generic parametrizations.  A fully quantitative assessment embedding our hybrid EoSs within the posterior distributions of Refs.~\cite{Gorda:2022lsk,Brandes:2023bob,Komoltsev:2024lcr} would delineate which region of the $(x_{\sigma\Delta},\,x_{\omega\Delta})$ coupling-constant space remains viable after incorporating all current multimessenger data and would test whether the posterior weight assigned to strong transitions increases when microphysically motivated correlations are built into the prior.

In summary, this work makes three contributions to the study of $\Delta$-isobar matter in neutron stars.  First, we provide a transparent microphysical interpretation of why the $\Delta$ onset can be discontinuous---tracing it to a self-amplifying scalar feedback whose strength is governed by the coupling difference $\Delta x$---and show that the resulting phase transition is reproduced by the SW4L parametrization in a coupling window consistent with earlier reports in other RMF frameworks.  Second, we present the first computation of the non-radial $g$-mode supported by a $\Delta$-induced density discontinuity, obtaining frequencies and damping times that place these modes within the sensitivity window of future gravitational-wave observations, particularly via the resonant tidal excitation mechanism studied in Refs.~\cite{counsell:2025imi,pereira:2025dti}.  Third, we demonstrate that the static and dynamical signatures of the $\Delta$-driven transition---the $M$--$R$ knee, reduced $\Lambda$, and $g$-mode frequencies---overlap quantitatively with those predicted for hadron--quark hybrid stars, sharpening the masquerade problem~\cite{Alford:2004hst} and complicating the interpretation of future detections.

Several directions for further work are worth highlighting. Embedding our hybrid EoSs within the posterior distributions of recent model-agnostic analyses~\cite{Gorda:2022lsk,Brandes:2023bob,Komoltsev:2024lcr} would delineate the viable region of the $(x_{\sigma\Delta},\,x_{\omega\Delta})$ parameter space and test whether the posterior weight assigned to strong transitions increases when microphysically motivated correlations are built into the prior.  Computing the tidal overlap integrals for the $\Delta$-induced $g$-mode, following the framework of Ref.~\cite{counsell:2025imi}, would yield direct predictions for the gravitational-wave phase shift during binary inspiral. Finally, combining the $g$-mode signature with percent-level radius measurements from future NICER-class missions and with thermal evolution calculations that distinguish $\Delta$-mediated neutrino channels from quark emissivity processes could, in principle, break the degeneracy between a ``$\Delta$-star'' and a quark-hybrid star.

\begin{acknowledgments}
M.O.C-P is a doctoral fellow of CONICET (Argentina). M.O.C-P, I.F.R-S and M.G.O acknowledge UNLP and CONICET (Argentina) for financial support under grants G187 and PIP 0169. GL acknowledges the financial support from the Brazilian agency CNPq (grant 316844/2021-7).
\end{acknowledgments}

\bibliography{biblio}

\end{document}